# Inhomogeneous cluster DMFT study on p-n junction Mott gap closing in Hubbard lattice

Florian Brette,[1] Eunjae Jeong,[1] Hongchul Choi,[2,3] Ji Hoon Shim,[3] Geunsik Lee[1]

[1]Department of Chemistry, Ulsan National Institute of Science and Technology, Ulsan 44919, Republic of Korea

[2]Max Planck POSTECH Center for Complex Phase of Materials, Pohang University of Science and Technology (POSTECH), Pohang 37673, Republic of Korea

[3]Department of Chemistry, Pohang University of Science and Technology (POSTECH), Pohang 37673, Republic of Korea

**Abstract**

Mott gap closing induced by an external potential gradient is studied by considering the Hubbard model on a square lattice. Our model consists of hole- and electron-doped domains forming the p-n junction where the on-site potential varies smoothly across the interface. The effect of the potential drop direction, either along square edge or diagonal, is studied by describing the correlation effects of inhomogeneous 2x2 plaquette impurity clusters within a dynamical mean-field theory framework. It is found that the metallic phase is favored over the Mott insulators when the potential drop becomes comparable to the energy eigenvalue difference of an open 2x2 plaquette between the ground state and the first excited state. Given the same associated electric-field strength, the excitation energy is almost independent of the drop direction, which amounts ~0.35t at U=5.8t with the hopping parameter t. Upon insulator-to-metal transition, the correlated metal exhibits a coherent peak at zero energy whose intensity is greater for the drop direction along the square diagonal than along the edge. The main origin is attributed to the preserved degeneracy of two molecular orbitals for the former case, while it undergoes splitting in the energy levels for the latter. Our result is discussed regarding the experimentally measured Mott gap closing thresholds.

## I. INTRODUCTION

Mott insulators are insulating materials that exhibit a correlation-induced energy gap due to repulsive Coulomb interaction between electrons. By controlling electronic parameters, they can be tailored to exhibit novel phenomena such as magnetism and superconductivity. A wide range of studies have been conducted to understand the microscopic mechanisms driving their phase transitions and to discover materials with desirable properties. Recently, external stimuli such as electric field or photon have been used to switch from an insulator to a metal. It possesses variety of potential applications in electronics, spintronics and sensing.[1] As a notable example, narrow-gap Mott insulators exhibiting the resistive switching[2] are promising for memristor devices.

The resistive switching behavior of Mott insulators remains controversial and material dependent. For instance, local heating has been shown to drive the transition in $VO_2$,[3] whereas a purely electronic origin has been suggested on phenomenogical grounds.[2,4,5] To unveil this microscopic origin, several studies have accounted for electron correlation effects by using the dynamical mean-field theory method (DMFT) with consideration of external leads.[6,7] However, the impurity models considered in these previous studies were limited to a single-site, leaving an incomplete understanding of non-local correlation effects. Accounting for these non-local correlation effects is essential for a qualitative and quantitative description of the physical properties.[8-10]

In this paper, the directional effect of an external potential gradient is studied by considering the short-range correlation effect. It is found that the metallic phase is favored over the Mott insulators when the potential drop becomes comparable to the excitation energy of an impurity from the ground-state and the first-excited state. While the threshold field strength is almost independent of the drop direction, the correlated metal shows stronger coherence for a gradient direction along the square diagonal than along the edge. The main origin is attributed to the preserved degeneracy of two molecular orbitals for the former case, while it undergoes splitting in the energy levels for the latter.

## II. METHOD

We consider two-dimensional Hubbard model with the on-site Coulomb repulsion $U$ and the hopping parameter $t$.

$$H = -t \sum_{<i,j>,\sigma} c_{j\sigma}^{\dagger} c_{i\sigma} + U \sum_i n_{i\uparrow} n_{i\downarrow} - \mu \sum_{i,\sigma} n_{i\sigma} + V^{ext} \quad (1)$$

Here, $c_{i\sigma}^{\dagger}$ ($c_{i\sigma}$) is the creation (annihilation) operator for an electron with spin $\sigma$ at site $i$, $n_{i\sigma} = c_{i\sigma}^{\dagger} c_{i\sigma}$ is the corresponding number operator. In this paper, the hopping parameter $t$ is chosen as the unit of energy. $V^{ext}$ is the external potential that varies periodically along one crystallographic direction, either [10] or [11], while remaining constant along the transverse direction. As shown in Figure 1, one supercell contains two homogeneous domains of hole-doped ($V^{ext} > 0$) and electron-doped ($V^{ext} < 0$), connected by each of two inhomogeneous transition regions. The potential profile in the inhomogeneous regions is defined by $\tanh(x/\lambda)$, with the empirical length scale $\lambda$ determining the steepness of the variation, and occupation is maintained at half-filling on average. Along the transverse direction, we employ a minimal periodic length with sampling 200 $k$-points. The chemical potential is fixed as $\mu = U/2$, with $U$ = 5.8 that is equal to the upper critical $U$ value for the metal-to-insulator transition in consistency with literatures.[8,9]

To describe the short-range correlation effects, we employ DMFT with multiple clusters indexed by $I$ along the longitudinal direction, i.e. [10] or [11]. Each cluster is a 2x2 plaquette consisting of four sites, such that every cluster is linked to four adjacent clusters via the nearest neighbor hopping. The impurity problem is solved using continuous time quantum Monte Carlo (CTQMC) simulations at inverse temperature β=40 and the impurity Green's function (GF) of $I$-th cluster is given by $G_I = [(iw + \mu - V_I^{ext})\mathbf{1} - \Sigma_I - \Delta_I]^{-1}$. Given the cluster self-energies $\Sigma_I(iw)$, we calculate the local cumulant matrices $M_I = [(iw + \mu)\mathbf{1} - \Sigma_I]^{-1}$ and periodize them by $M_k = \sum_{i,j} M_{ij} e^{i\mathbf{k}\cdot(\mathbf{r}_i - \mathbf{r}_j)}$ to construct the lattice GF $G_k = [M_k^{-1} - h_k]^{-1}$ for the transverse normal mode $k$, where $h_k$ encompasses the lattice hopping and the external potential in Equation (1). The local GF is obtained by $G_{loc} = \int G_k \, dk$. The DMFT self-consistency condition, $\hat{P}_I G_{loc} = G_I$, is enforced for every cluster $I$, with $\hat{P}_I$ denoting the projection operator onto the $I$-th cluster. The hybridization matrix $\Delta_I$ is diagonalized using the Hartree-Fock self-energy $\Sigma_I(iw = \infty)$. Finally, spins are symmetric in our calculations.

Figure 2 illustrates the arrangement of independent clusters whose self-energies are updated during the DMFT cycle. The two cases correspond to $V^{ext}$ varying along the [10] direction in Fig. 2(a) or [11] in Fig. 2(b), where one supercell contains 50 or 60 clusters, respectively. For [10], the left and right homogeneous regions consist of 20 clusters each, and the two inhomogeneous regions contain 5 clusters each. Similarly, for the [11] direction, there are 23 and 7 clusters for the homogeneous and the inhomogeneous regions, respectively. The number of clusters used for the homogeneous regions was sufficient to have consistency with the solutions of independent DMFT calculations for homogeneous bulk. The value of $V^{ext}$ is fixed as +0.5 or -0.5 for hole- or electron-doped region, respectively, and the cluster self-energies are constrained to be equal to that of the homogeneous DMFT calculation, which are denoted by L or R bulk reservoir in Figure 2. For each of [10] and [11] directions, the two inhomogeneous regions are related by inversion symmetry, thus equivalent clusters are provided with the same cluster index in Figure 2. We use enough clusters in the inhomogeneous regions such that the cluster next to the homogeneous region has a self-energy reasonably close to L or R. To investigate the effect of the gradient direction on the metallic transition, the DMFT self-consistency cycles are initialized with Mott-insulating self-energies, with each iteration requiring an independent CTQMC calculation for every unique cluster.

## III. RESULTS AND DISCUSSION

Our correlated system is solved for different steepness of $V^{ext}$ variation. Upon reaching self-consistency, only the central cluster ($I$=0 in Figure 2) remains Mott insulator type when the steepness is below a critical threshold (see Figures S1 and S2). Beyond this threshold, however, it changes to metal state. For example, Fig. 3 plots the on-site potential values for the clusters within the inhomogeneous region. Figure 3(a) shows five clusters along [10] with the dashed lines denoting the inter-cluster hoppings, while Figure 3(c) shows seven clusters along [11]. In the [11] case, the clusters are separated by the nearest neighbor hoppings as in Figure 2(b). The central cluster with *I*=0 (or having $x_{10} = 0$, $x_{11} = 0$) has two positive and two negative values of equal magnitudes for [10] as in Fig. 3(a), or one positive and one negative values of equal magnitude with two zero's for [11] as in Fig. 3(c). The magnitude of non-zero values, denoted by dV, indicates the potential steepness. Figure 3(b) presents the imaginary part of the Matsubara self-energy for the two dV values from Fig. 3(a). The insulator-to-metal transition is found to occur between dV=0.23 and dV=0.28 in the [10] case. Similarly, Fig. 3(d) shows the results of [11] case with the transition occurring between dV=0.24 and dV=0.27.

From Figs. 3(b) and 3(d), the critical dV value where the Mott insulator changes to a metal is found to be similar between [10] and [11] cases. By taking the mean, the predicted values are 0.254 for the former and 0.255 for the latter. However, the associated field strength is significantly smaller for [11] than [10], because the diagonal of a 2x2 plaquette is $\sqrt{2}$ times longer than its edge. We suggest that this is due to enhanced hybridization when the impurity sites contain a site with zero on-site level. To verify this, we modified the [10] model to include a zero-potential level for the central cluster, which requires an asymmetric setup as shown in Fig. 4(a) (see Fig. S3 for asymmetric arrangement of the clusters in the unit cell). The modified [10] model is denoted as [10]-Asym hereafter. For [10]-Asym, we observed the transition between dV=0.15 and dV=0.17, which is significantly lower than that (dV~0.25) by the original [10] model, and confirm the enhancement of the hybridization in the presence of zero on-site level .

The insulator-to-metal transition is analyzed via the imaginary part of the central (*I*=0) impurity GF at the lowest Matsubara frequency, that is $\mathrm{Tr}\{\mathrm{Im}[G_{I=0}(iw_1)]\}$. The variation of $V^{ext}$ per longitudinal distance (the edge length equals to one) within the central cluster is used to describe the strength of a hypothetical electric field, denoted by $\varepsilon^{ext}$. It is equal to 2 dV and $\sqrt{2}$ dV for [10] and [11], respectively. Figure 4(b) plots the results for the three models, [10], [11], and [10]-Asym, where we varied the $\lambda$ values manually near the transition. An abrupt change of $\mathrm{Im}[G_{I=0}(iw_1)]$ is found for two adjacent values of $\varepsilon^{ext}$. It indicates a signature of transition to a correlated metal. Such transition occurs at $\varepsilon^{ext}\sim 0.35$ by the [11] and [10]-Asym models, while at much higher (~0.5) by the [10] model. Hence, it is confirmed that the presence of zero on-site potential level and the resulting enhanced hybridization is a main factor for lowering the transition threshold. Although regarding the former two models the transition by [10]-Asym occurs at a slightly lower value than that by [11], its validity is uncertain because of limited precision of our approach to identify the transition point.

Figure 4(b) shows that the metal states exhibit higher intensity of $\mathrm{Tr}\{\mathrm{Im}[G_{I=0}(iw_1)]\}$ for [11] case than [10]-Asym. Consistently, Figs. 3(b) and 3(d) show that the scattering rate (or imaginary part of self-energy at $iw\rightarrow 0$) is relatively low for [11]. Related to this, spectra of the centeral cluster, $A_{I=0}(w)$, were calculated using the maximum entropy method for analytic continuation. For each of [10]-Asym and [11] models, one configuration for insulator and two for metal were selected, which are denoted by (i)-(iii) or (iv)-(vi) in Fig. 4(b). Figures 4(c) and 4(d) plot the spectra for [10]-Asym and [11] cases, respectively. For the insulator phases, shown by the black solid line, narrow peaks appear at a lower edge of the conduction band and at the higher edge of the valence band. They are signatures of localized holon-doublon pairs as discussed in literatures.[11,12] In the correlated metal phases, shown by the red and blue solid lines, the metallic spectra at w=0 exhibits a double-peak structure. However, the peak intensity is greater for [11] than [10]-Asym, and a dip feature is observable for [10]-Asym. This suggests more coherent behavior for [11] than [10] (see Fig. S4 for the symmtric [10]). Thus, the quasiparticle coherence is found to depend on direction of $\varepsilon^{ext}$.

Tendency of quasiparticle coherence against $\varepsilon^{ext}$ direction is studied by considering an isolated 2x2 plaquette. In Figure 5(a), the single-electron energy level, $E^{N=1}(\varepsilon^{ext})$, is plotted for four molecular orbitals as a function of $\varepsilon^{ext}$ for [10] and [11] directions. When $\varepsilon^{ext} = 0$, there are two degenerate orbitals at E=0. These orbitals have the symmetry of (+0.5, -0.5, +0.5, -0.5) and (+0.5, +0.5, -0.5, -0.5), or their linear combinations, following site-index order of bottom-left, bottom-right, top-left, top-right. For $\varepsilon^{ext} \neq 0$, the orbitals near E=0 exhibit energy splitting for [10] direction, while they remain degenerate for [11]. In other words, the crystal potential varying along [10] lifts degeneracy of the two degenerate molecular orbitals, but the one along [11] preserves it.

In Figure 5(b), the energy of four-electron system, $E^{N=4}(\varepsilon^{ext})$, by exact diagonalization is plotted for two lowest energy states as a function of $\varepsilon^{ext}$. We find no significant difference between two directions, i.e. almost overlapping solid-black and dashed-red lines, in the plotted range. It is found that the ground state is a singlet with S=0, i.e. |N=4,S=0>, and the first-excited state is a triplet with S=1, i.e. |N=4,S=1>. The calculated excitation energy is ~0.35 without significant variation about $\varepsilon^{ext}$. This value is comparable to the threshold value of $\varepsilon^{ext}$, shown in Fig. 4(b). This is can be understood that the energy bands associated with the Weiss field are broadened by the on-site potential splitting (+dV and –dV), and that the increased kinetic energy makes the first excited state accessible thermodynamically. Consistently, as provided in the Table 1, our CTQMC calculations show that the sampling probabilities for the S=0 and S=1 states varied significantly across the insulator-to-metal transition.

Regarding the observed difference of coherency between [10] and [11], Fig. 5(c) represents the occupation of four molecular orbitals associated with $E^{N=1}(\varepsilon^{ext} \neq 0)$ for ground state |N=4,S=0> and first excited state |N=4,S=1>. For [10] direction, excitation from S=0 to S=1 could involve an electron to be scattering into a different energy level. In contrast, the [11] case would not have such scattering, which could be the main reason for enhanced coherence.

The effect of increasing the cluster size beyond current 2x2 is discussed. From several studies on the half-filled 2D Hubbard model using cluster or cellular DMFT,[9,13] the extended antiferromagnetic correlation reduces the critical U value for the metal-to-insulator crossover. In Table 2, we list the calculated energy eigenvalues of the ground S=0 and the first-excited S=1 states for an open half-filled cluster whose size varies as 2x2, 2x4, and 4x4, using U=5.8. It shows that the excitation energy, written in the parenthesis, descrease as the cluster size increase. It is also known that such excitation energy for strong coupling limit approximately equals to $4t^2/U$. Our DMFT result suggests that the insulator-to-metal transition would occur when the on-site potential difference exceeds the excitation energy from S=0 and S=1. Based on this, the threshold field strength is estimated as 0.342/1 ~ 0.3 for 2x2, 0.179/3 ~ 0.06 for 4x4. It can be said that the extended correlation beyond 4x4 could reduce the thrshold field strength significantly.

Experimentally, the dielectric breakdown for narrow-gap Mott insulators occurs at $\sim 10^4$ V/cm.[2] Considering the Zener mechanism, an on-site Coulomb repulsion ~3 eV and an intersite distance ~0.3 nm yield a predicted threshold field of $\sim 10^8$ V/cm, which overestimates the experimental value by four orders of magnitude. Using our result of the 4x4 cluster, a predicted threshold value is ~0.06 V / 0.3 nm = $2\times10^6$ V/cm, assuming the hopping magnitude t = 1 eV. It is still two orders of magnitude greater than the experimental value. Nevertheless, our results indicates an important role of non-local correlation on the Mott gap closing, and theoretical prediction could be improved if one extends the correlation length and considers material specific parameters.

## IV. CONCLUSIONS

In summary, the direction effect of the external field on the Mott gap closing has been studied, using cluster DMFT on an inhomogeneous 2x2 plaquette system. The insulator to metal transition occurs when the field strength is comparable to the excitation energy ~0.35 of the impurity. The threshold values for two studied directions ([10] and [11]) are considered equal within precision of our model calculations. Beyond the transition, the higher coherence is observed for [11] and it is attributed to the single-electron orbital behavior: the associated molecular orbitals exhibit a lifted degeneracy for [10] field direction, causing incoherent scattering, whereas the orbitals remain degenerate for [11]. Our study indicates an important role of non-local correlation on the Mott gap closing.

## ACKNOWLEDGMENTS

This work was supported by the National Research Foundation of Korea (NRF) grant funded by the Korean government (Grants No. RS-2025-00522876). H. Choi was supported by the National Research Foundation of Korea (NRF) funded by the Ministry of Science and ICT (No. RS-2022-NR068223). Computation resources were supported by KISTI (Grants No. KSC-2025-CRE-0372, No. KSC-2024-CRE-0482). We thank the supercomputing resources of the UNIST Supercomputing Center. This work was supported by the InnoCORE program of the Ministry of Science and ICT(1.260005.01).

**REFERENCES**

[1] A. Milloch, M. Fabrizio, C. Giannetti, npj Spintronics **2**, 49 (2024).

[2] P. Stoliar, L. Cario, E. Janod, B. Corraze, C. Guillot-Deudon, S. Salmon-Bourmand, V. Guiot, J. Tranchant, M. Rozenberg, Adv. Mater. **25**, 3222 (2013).

[3] A. Zimmers, L. Aigouy, M. Mortier, A. Sharoni, S. Wang, K. G. West, J. G. Ramirez, I. K. Schuller, Phys. Rev. Lett. **110**, 056601 (2013).

[4] P. Diener, E. Janod, B. Corraze, M. Querré, C. Adda1, M. Guilloux-Viry, S. Cordier, A. Camjayi, M. Rozenberg, M. P. Besland, L. Cario, Phys. Rev. Lett. **121**, 016601 (2018).

[5] V. Guiot, L. Cario, E. Janod, B. Corraze, V. Ta Phuoc, M. Rozenberg, P. Stoliar, T. Cren, D. Roditchev, Nature Commun. **4**, 1722 (2013).

[6] H. Aoki, N. Tsuji, M. Eckstein, M. Kollar, T. Oka, P. Werner, Rev. Mod. Phys. **86**, 779 (2014).

[7] J. Li, C. Aron, G. Kotliar, J. E. Han, Nano Lett. **17**, 2994 (2017).

[8] H. Park, K. Haule, G. Kotliar, Phys. Rev. Lett. **101**, 186403 (2008).

[9] T. Schäfer, F. Geles, D. Rost, G. Rohringer, E. Arrigoni, K. Held, N. Blümer, M. Aichhorn, A. Toschi, Phys. Rev. B **91**, 125109 (2015).

[10] H. Terletska, S. Iskakov, T. Maier, E. Gull, Phys. Rev. B **104**, 085129 (2021).

[11] S. Zhou, Y. Wang, Z. Wang, Phys. Rev. B **89**, 195119 (2014).

[12] S.-S. B. Lee, J. von Delft, A. Weichselbaum, Phys. Rev. Lett. **119**, 236402 (2017).

[13] M. Meixner, H. Menke, M. Klett, S. Heinzelmann, S. Andergassen, P. Hansmann, T. Schäfer, SciPost Phys. **16**, 059 (2024).

**Table 1.** Sampling probabilities of the ground ($P_{S=0}$) and the first-excited ($P_{S=1}$) states in four-electron subspace. The used impurity solver is CTQMC. The values of $\varepsilon^{ext}$ for the two adjacent points associated with the insulator-to-metal transtion are given in the second column along with the indicators in Figure 4(a), (i) and (ii) for [10]-Asym, (iv) and (v) for [11].

| Model | $\varepsilon^{ext}$ | $P_{S=0}$ | $P_{S=1}$ |
|---|---|---|---|
| [10]-Asym | (i) 0.304 | 0.573 | 0.077 |
| | (ii) 0.341 | 0.364 | 0.177 |
| [11] | (iv) 0.341 | 0.584 | 0.071 |
| | (v) 0.381 | 0.317 | 0.197 |

**Table 2.** Calculated two lowest energy eigenvalues (ground state S=0, first excited state S=1) of a half-filled, open two-dimensional Hubbard cluster using exact diagonalization. The right three columns correspond to the cluster sizes (2×2, 2×4, and 4×4), and the used on-site Coulomb repulsion is U=5.8 that lead to the insulating phase near the metal-insulator crossover for 2x2. The values in the parenthesis are the excitation energy from S=0 to S=1. The energy unit is the nearest neighbor hopping.

| Spin | 2x2 | 2x4 | 4x4 |
|---|---|---|---|
| S=0 | -1.673 | -3.886 | -8.759 |
| S=1 | -1.331 (0.342) | -3.607 (0.279) | -8.580 (0.179) |

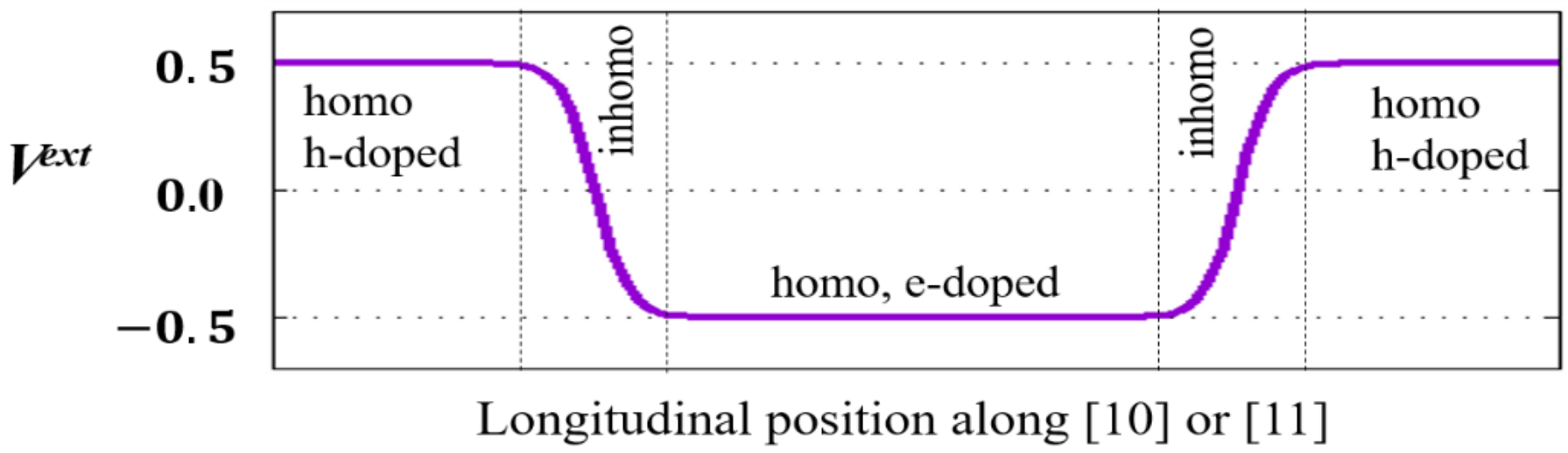

**Figure 1**. The profile of external potential, $V^{ext}$, in the inhomogeneous (inhomo in the figure) region between homogeneous (homo) hole-doped (h-doped) and electron-doped (e-doped) domains. By the periodic boundary condition, one supercell contains two inhomogeneous regions or two junctions. The considered direction across the junction is [10] and [11] in the square lattice.

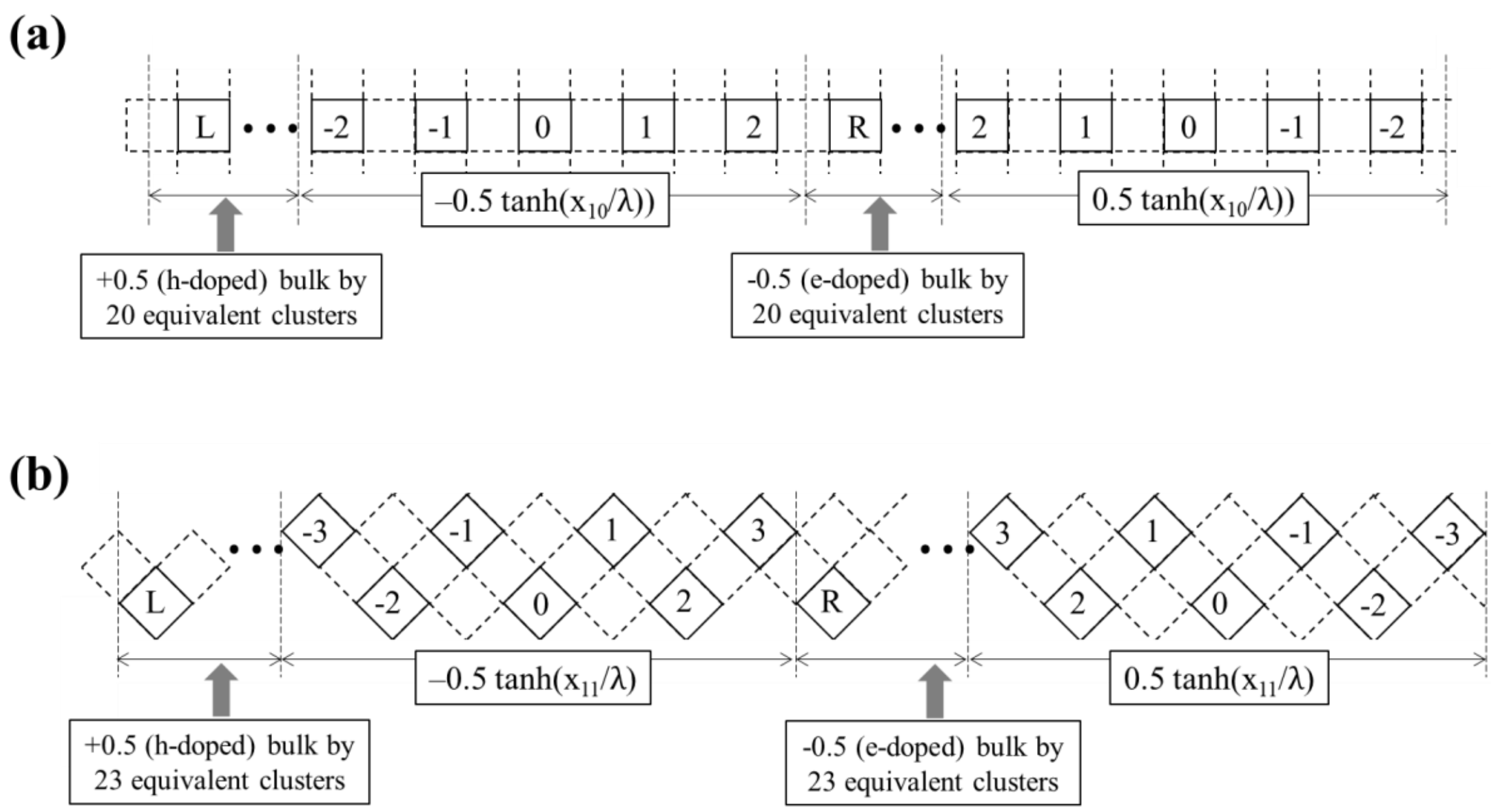

**Figure 2**. Indices of inequivalent impurity clusters per one supercell when the external potential is varied along (a) [10], (b) [11]. For [10] or [11] case, there are total 50 or 60 clusters of 2x2 plaquette type, respectively. Owing to the inversion symmetry, the inhomogeneous region between hole-doped and electron-doped regions contains only five independent clusters indexed by $I$ = -2, -1, 0, 1, 2 for the former case, while seven (-3, -2, -1, 0, 1, 2, 3) for the latter case. The potential profile is defined by tanh(x/λ) with x($x_{10}$ or $x_{11}$) = 0 corresponding to the most neutral site.

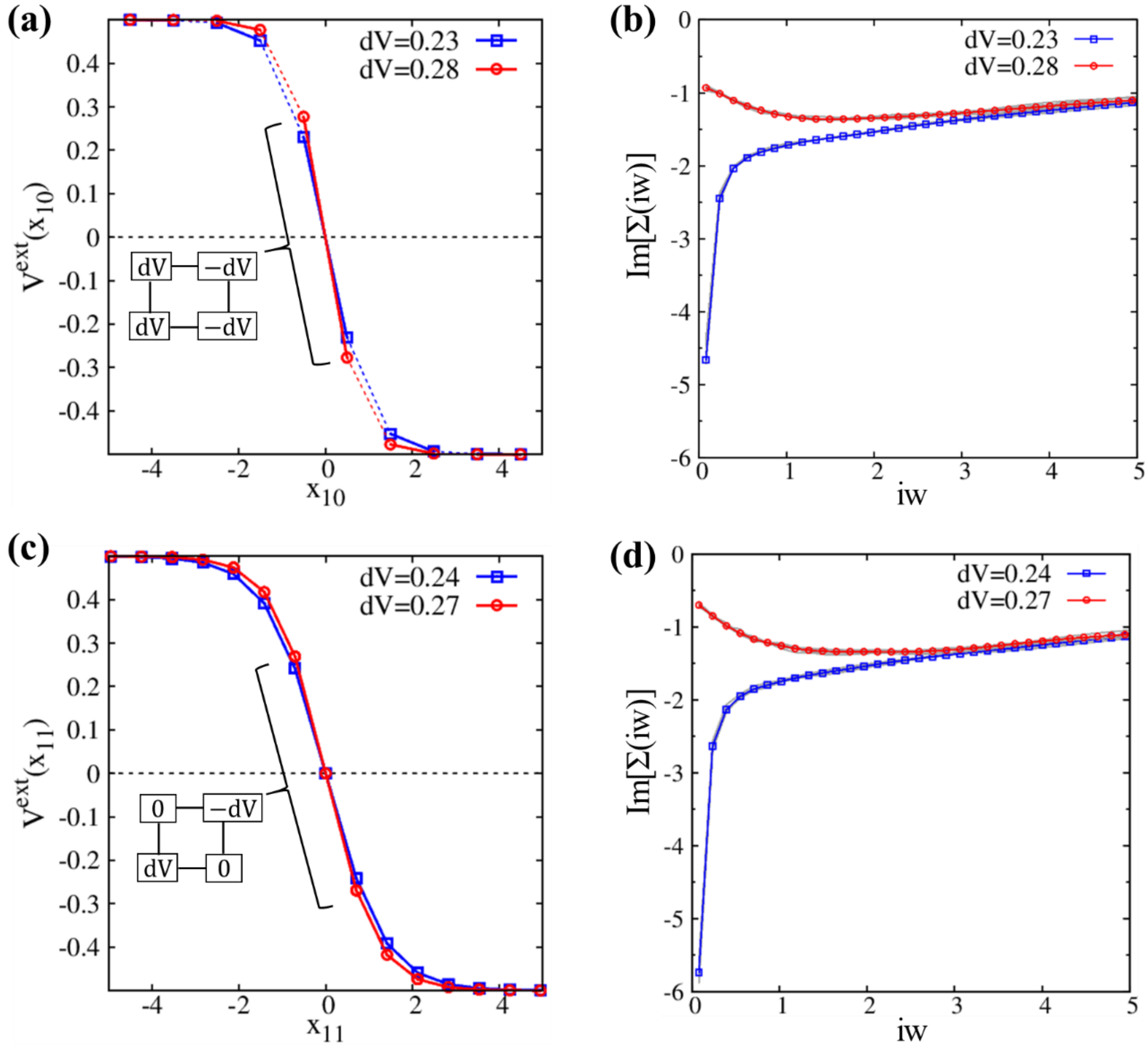

**Figure 3**. Plot of $V^{ext}$ values in the left inhomogeneous region in Figure 2 for (a) [10] and (c) [11] cases. Two profiles are plotted for each case, where the steepness is given by the magnitude of dV at the central cluster having $x_{10}$=0 or $x_{11}$=0. (b) and (d) are the imaginary part of Matsubara self-energy of the central cluster for the $V^{ext}$ profiles along [10] and [11], respectively.

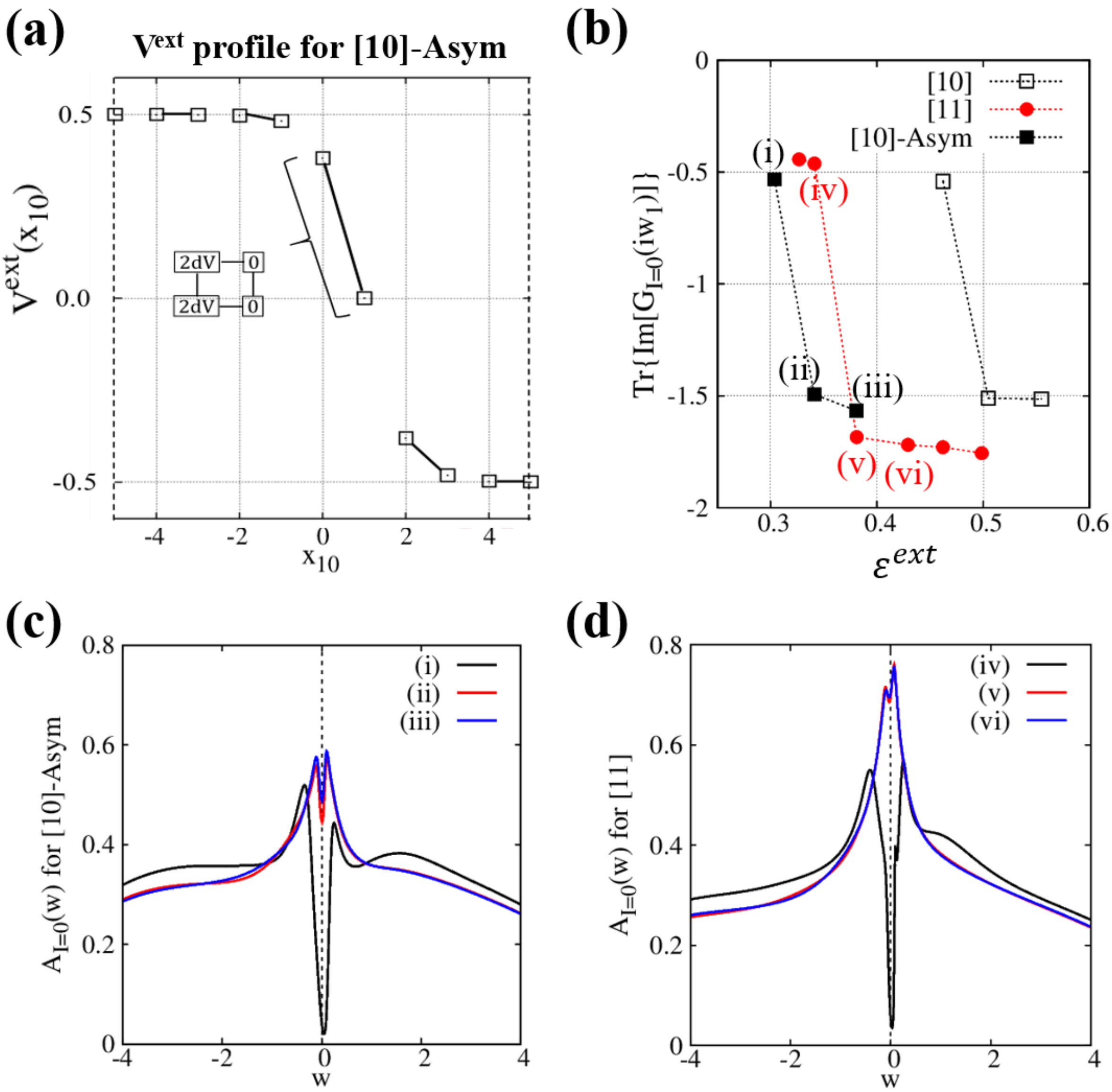

**Figure 4**. (a) Profile of $V^{ext}$ for [10]-Asym case, which is similar to Fig. 3(a) but the zero potential level site included. (b) Imaginary part of the central impurity Green's function at the lowest Matsubara point for chosen values of the field strength ($\varepsilon^{ext}$) along [10] and [11] directions. (c) and (d) are the spectra calculated for three selected values of $\varepsilon^{ext}$, (i)-(iii) for [10]-Asym or (iv)-(vi) for [11] in (b).

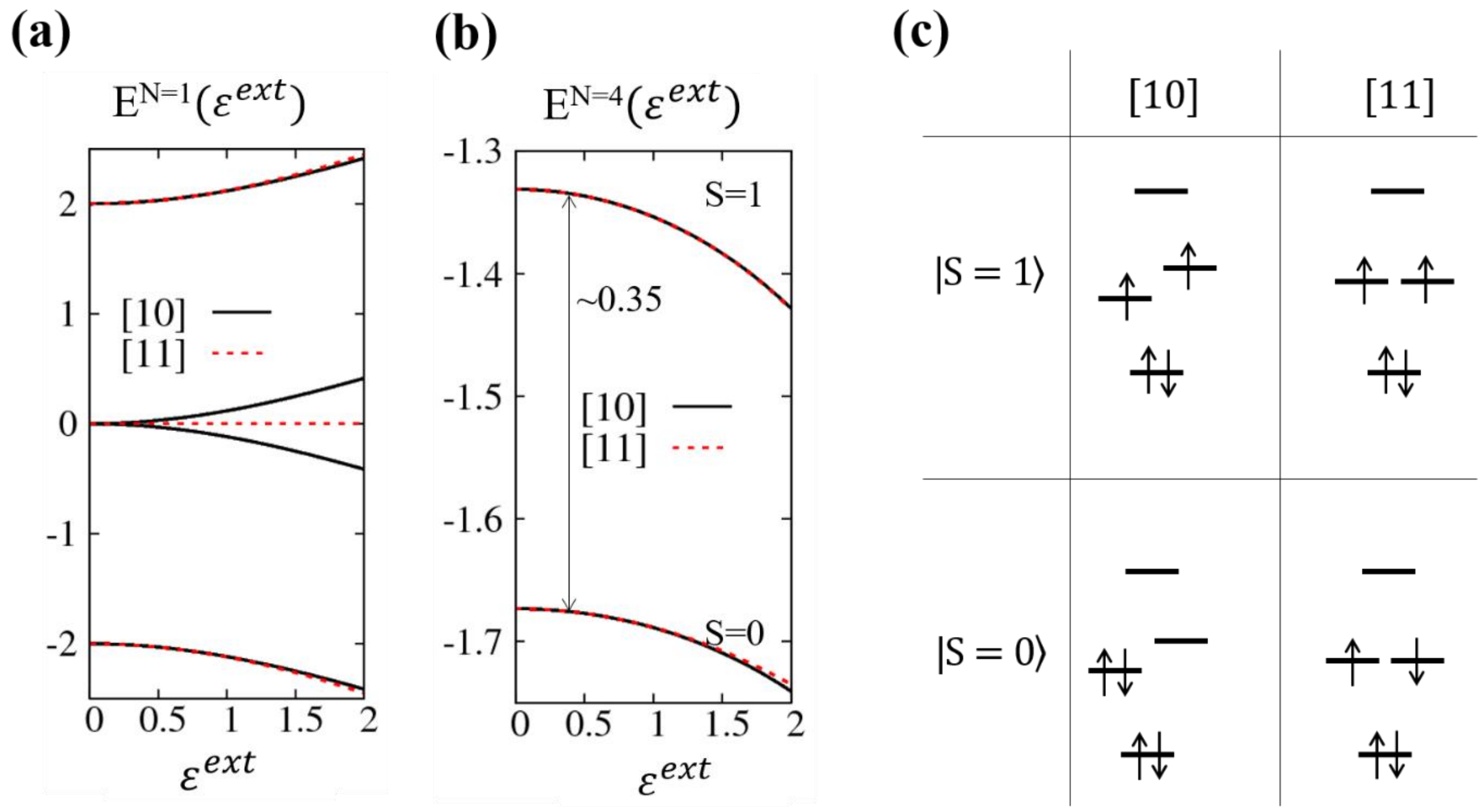

**Figure 5**. (a) Energy levels of four molecular (or single-electron) orbitals and (b) two lowest energy eigenvalues (ground state S=0 and first excited S=1) of four-electron states by exact diagonalization for an isolated 2x2 plaquette against $\varepsilon^{ext}$ along [10] and [11] directions. (c) Occupation representation of the four-electron ground state |S=0> and first excited state |S=1> for non-zero $\varepsilon^{ext}$ along [10] and [11] directions. The four horizontal bars indicate the molecular orbital energy levels as plotted in (a). The up and down arrows denote +0.5 and -0.5 values of Sz, respectively.